\documentclass[a4paper,10pt,conference]{ieeeconf}

\IEEEoverridecommandlockouts
\usepackage{graphicx,psfrag,enumerate}
\usepackage{amssymb}
\usepackage{amsmath}
\usepackage{calrsfs}
\usepackage{times,bm}
\usepackage{subeqnarray}
\usepackage{latexsym,psfrag}

\usepackage{subfigure}

\newtheorem{definition}{Definition}
\newtheorem{theorem}{Theorem}
\newtheorem{remark}{Remark}

\newtheorem{lemma}{Lemma} 
\newtheorem{corollary}{Corollary} 
\newtheorem{proposition}{Proposition}

\def\diag{\mathop{\mathrm{diag}}\nolimits} % Diagonal matrix
\def\Ker{\mathop{\mathrm{Ker}}\nolimits} % The null-space of a matrix
\begin{document}

\title{Detectability of distributed
  consensus-based observer networks:\\ An elementary analysis and
  extensions\thanks{This research was supported under Australian Research 
    Council's Discovery
Projects funding scheme (Project number DP120102152).}}

\author{V. Ugrinovskii\thanks{School of Engineering and IT, University of
    New South Wales Canberra
  at the Australian Defence Force Academy, Canberra, Australia, Email:
  v.ugrinovskii@gmail.com}} 

\maketitle
          
% \begin{keyword}                           % Five to ten keywords,  
% Large-scale systems, distributed filtering, robust filtering, consensus,
% vector dissipativity, vector Lyapunov functions, Markovian jump systems.
% \end{keyword}

\begin{abstract}
This paper continues the study of local detectability and
observability requirements on components of distributed observers networks 
to ensure detectability properties of the network. First, we present a
sketch of an
elementary proof of the known result equating the
multiplicity of the zero eigenvalue of the Laplace matrix of a digraph
to the number of its maximal reachable subgraphs.
Unlike the existing algebraic proof, we use a direct analysis
of the graph topology. This result is then used in the second part of the
paper to extend our previous results which connect the detectability of an
observer network with corresponding local detectability and observability
properties of its node observers. The proposed extension allows for nonidentical
matrices to be used in the interconnections.  
\end{abstract}

\section{Introduction}
The principle of distributed estimation can be traced back to the original work
on decentralized estimation completed in the
90s~\cite{BD-1992,CS-1995}, while the more modern ideas are focused around
the observer network design to allow node
estimators to exchange information with the objective of improving their
knowledge of the system 
state (or a part thereof) through reaching an agreement about the estimated
quantity. 
A most recent development in this area is
concerned with  $H_\infty$ distributed estimation in the presence of modelling
uncertainties and perturbations~\cite{SWH-2010,U6,LaU1,U7,UFri1}.      

Some efforts have recently been made to obtain
a more detailed insight into the role played by the communication topology
in distributed estimation problems. For example, in~\cite{DK-2013} an
algebraic design of the communication topology for networked observers was
considered using the structured systems theory. The objective in
\cite{EM-2013} was to maintain
the collective detectability of the network while achieving a desired
observer sparsity. Here, the term \emph{collective detectability}, or
\emph{distributed detectability}, refers to the detectability property
achieved by the entire network, in contrast to the \emph{local
  detectability} which refers to the detectability of the plant from the
measurements taken by individual nodes of the network. In~\cite{U7b-journal} the
property of distributed detectability of the observer network was related
to the local detectability of the plant through measurements, observability of
the node filters through interconnections, and the largest spanning trees of the
underlying communication graph. In particular, it was shown that for the
associated interconnected system of filtering error dynamics to be
stabilizable via output injection, each network component spanned by a
non-extendable tree must be collectively detectable. A similar requirement
on the nodes with unobservable local error dynamics to have an 
incoming path connecting them with collectively observable subnetworks
was discussed in~\cite{FFS-2010}.  

% Reference~\cite{FFS-2010} also discussed the requirement
% on the network topology in relation to achieving a desired pole 
% assignment for the error dynamics arising in the distributed moving average
% estimator. Specifically, the conditions in~\cite{FFS-2010} require each
% node with unobservable local error dynamics to have an incoming path
% connecting it with a set of observable nodes to allow the information from
% observable network nodes to be shared with those nodes which have
% inadequate sensors. More generally, in~\cite{U7b-journal} the
% property of distributed detectability of the observer network was related
% to local detectability of the plant through measurements, observability of
% the node filters through interconnections, and spanning trees of the
% underlying communication graph. It was shown that for the interconnected
% filtering error dynamics to be stabilizable via output injection, the
% network components spanned by trees must be collectively detectable
% (\emph{regional detectability} in the language of~\cite{FFS-2010}). 

In this paper, we revisit the distributed detectability problem for a
network of interconnected state estimators observing a linear plant
considered in~\cite{U7b-journal}. The detectability conditions
obtained in that reference are based on a relationship between the
multiplicity of the zero eigenvalue of the graph Laplacian matrix and the
number of maximal reachable clusters within the
graph~\cite{AC-2005,CV-2006}. In these references, the mentioned
relationship was obtained as a special case of a more general
theory, using the tools from the matrix algebra. Here we give an elementary
self-contained proof of this relationship using a direct analysis of the
graph topology. Furthermore, we use this result to present some extensions
of the results in~\cite{U7b-journal} which show that in 
the distributed estimation scenario, algebraic properties of the graph
Laplacian must be complemented by the detectability and observability
properties of the node filters through measurements and interconnections,
respectively. The proposed extensions relax one of the limitations of these
results in that we do not require all observers to use the
same matrix for interconnections.  

The paper is organized as follows. In Section~\ref{Kirchhoff} we provide
the alternative proof of the above mentioned result about the relationship
between the multiplicity of the zero eigenvalue of the Laplace matrix of
a directed graph and the number of its maximal reachable components. In
Section~\ref{main} this result is  
applied to establish the connection between collective detectability of the
plant and its detectability by individual sensor nodes combined into certain
clusters spanned by trees. This provides a natural
way to analyse the collective detectability properties of the entire filter
network from the corresponding properties of the nodes and
interconnections within clusters, using the results in~\cite{U7b-journal}.

\paragraph*{Notation} Throughout the paper, $\mathbf{R}^n$ denotes the real
Euclidean $n$-dimensional vector space, with the norm $\|x\|\triangleq
(x'x)^{1/2}$; here the symbol $'$ denotes the transpose of a matrix or a
vector. $\Ker A$ denotes the null-space of a matrix $A$. 
$\mathbf{0}_k\triangleq [0~\ldots~0]'\in \mathbf{R}^k$, 
$\mathbf{1}_k\triangleq [1~\ldots~1]'\in \mathbf{R}^k$, and
$I_k$ is the identity matrix; we will omit the
subscript $k$ when this causes no ambiguity. 
The symbol $\otimes$
  denotes the Kronecker product of matrices, or the tensor product of two
  vector spaces. Also we use the notation $\prod_{l=1}^N\mathcal{P}_l$ to
  denote the Cartesian product of $N$ vector spaces $\mathcal{P}_1,
  \ldots,  \mathcal{P}_N$. $\dim\mathcal{X}$ denotes the dimension of a
  finite dimensional vector space $\mathcal{X}$. The symbol $\diag[P_1,\ldots,P_N]$
  denotes the block-diagonal matrix, whose  diagonal
  blocks are $P_1,\ldots,P_N$.     

\section{The multiplicity of the zero eigenvalue of the digraph Laplacian is
  equal to the number of maximal subgraphs spanned by trees}\label{Kirchhoff}
 
Consider a filter network with 
$N$ nodes and a directed graph topology $\mathbf{G} = (\mathbf{V},\mathbf{E})$;
$\mathbf{V}=\{1,2,\ldots,N\}$ and $\mathbf{E}\subseteq \mathbf{V}\times \mathbf{V}$ are the set of vertices and the
set of directed edges, 
respectively. The ordered pair $(j,i)$ will denote the directed edge
of the graph originating at node $j$ and ending at node $i$. In accordance
with the common convention, the graph $\mathbf{G}$ is assumed to have no
self-loops, i.e., 
$(i,i)\not\in \mathbf{E}$. The notation $\mathbf{G}(i_1,\ldots,i_M)$ will
denote a subgraph of $\mathbf{G}$ with the node set
$\{i_1,\ldots,i_M\}\subseteq \mathbf{V}$ and an edge set
$\{(i_s,i_l):i_s,i_l\in \{i_1,\ldots,i_M\}\}\subseteq \mathbf{E}$.    

For each $i\in \mathbf{V}$, let $\mathbf{V}_i=\{j:(j,i)\in \mathbf{E}\}$
be the neighbourhood of $i$.  The cardinality of
$\mathbf{V}_i$, known as the in-degree of node $i$, is denoted $p_i$; i.e.,
$p_i$ is equal to the number of incoming edges for node $i$. 

Node $i$ of a digraph is said to be reachable from node $j$ if there exists
a directed path originating at $j$ and ending at $i$. The graph is
connected if any two nodes are connected by an undirected path; the graph is
strongly connected if its every node is reachable from any other node. A
subgraph of $\mathbf{G}$ is a spanning tree if it has the same vertex set
$\mathbf{V}$, has no cycles, has $N-1$
edges and 
contains a node from which every other node of $\mathbf{G}$ can be reached
by traversing along the directed edges of $\mathbf{G}$ (the root node). 

Throughout the paper, $\mathcal{A}$, $\mathcal{D}$ and $\mathcal{L}$ will denote
the adjacency matrix, the in-degree matrix and the Laplacian matrix of the
graph $\mathbf{G}$, respectively,
\begin{eqnarray*}
  && \mathcal{A}=\left[\mathbf{a}_{ij}\right]_{ij=1,\ldots, N}, \quad
  \mathbf{a}_{ij}= \begin{cases}1 & \text{if $(j,i)\in\mathbf{E}$}, \\
                                0 & \text{otherwise},
                              \end{cases} \\
  && \mathcal{D}=\diag{[p_1~ \ldots~ p_N]}, \\
  && \mathcal{L}=\mathcal{D}-\mathcal{A}.
\end{eqnarray*}
% whose elements are
% \begin{eqnarray}
% [\mathcal{L}]_{ij}=\begin{cases}
% p_i & j=i, \\
% -1 & j\in \mathbf{V}_i, \\
% 0, & j\not\in \mathbf{V},~j\neq i. 
% \end{cases}
% \end{eqnarray}
The eigenvalues of $\mathcal{L}$ will be denoted $\lambda_i$,
$i=1,\ldots,N$. It is easy to check that zero is one of the eigenvalues of
$\mathcal{L}$, and $\mathbf{1}_N$ is the corresponding right eigenvector.

% As is well known, this eigenvalue has multiplicity one if 
% and only if the interconnection graph has a spanning tree~\cite{RB-2005}. 
% Furthermore, the classical result in the graph theory states that the
% multiplicity of the zero eigenvalue of the Laplace matrix of an undirected
% graph is equal to the number of connected components of the
% graph. Recently, this result was extended to directed
% graphs~\cite{AC-2005,CV-2006}. It was shown in these references that the 
% multiplicity of the zero eigenvalue of $\mathcal{L}$ is equal to the number of
% maximal reachable sets within the graph. In~\cite{AC-2005,CV-2006} this
% result was obtained as a special case of a more general theory covering
% more general matrices, using algebraic tools. Here we present a direct
% proof of this result of~\cite{AC-2005,CV-2006}, based on the analysis of
% graph topology. Our proof is based on the observation that maximal
% reachable sets, termed in~\cite{CV-2006} as reaches, are
% spanned by a tree. The following definition makes this property explicit.   

\begin{definition}\label{cluster}
A subgraph of $\mathbf{G}$ is a \emph{cluster}, if it satisfies the following
requirements:
\begin{enumerate}[(i)]
\item
It contains a spanning tree; and
\item   
It is a maximal subgraph in the sense that none of its spanning trees can
be extended by adding nodes from the set $\mathbf{V}$. 
\end{enumerate}   
\end{definition}

\begin{figure}[t]
\mbox{}
\vspace*{10pt}

\psfrag{1}{1}
\psfrag{2}{2}
\psfrag{3}{3}
\psfrag{4}{4}
\psfrag{5}{5}
\psfrag{6}{6}
\psfrag{7}{7}
\psfrag{Cluster 1}{Cluster 1}
\psfrag{Cluster 2}{Cluster 2}
 \centering
 \subfigure[]{\includegraphics[width=\columnwidth]{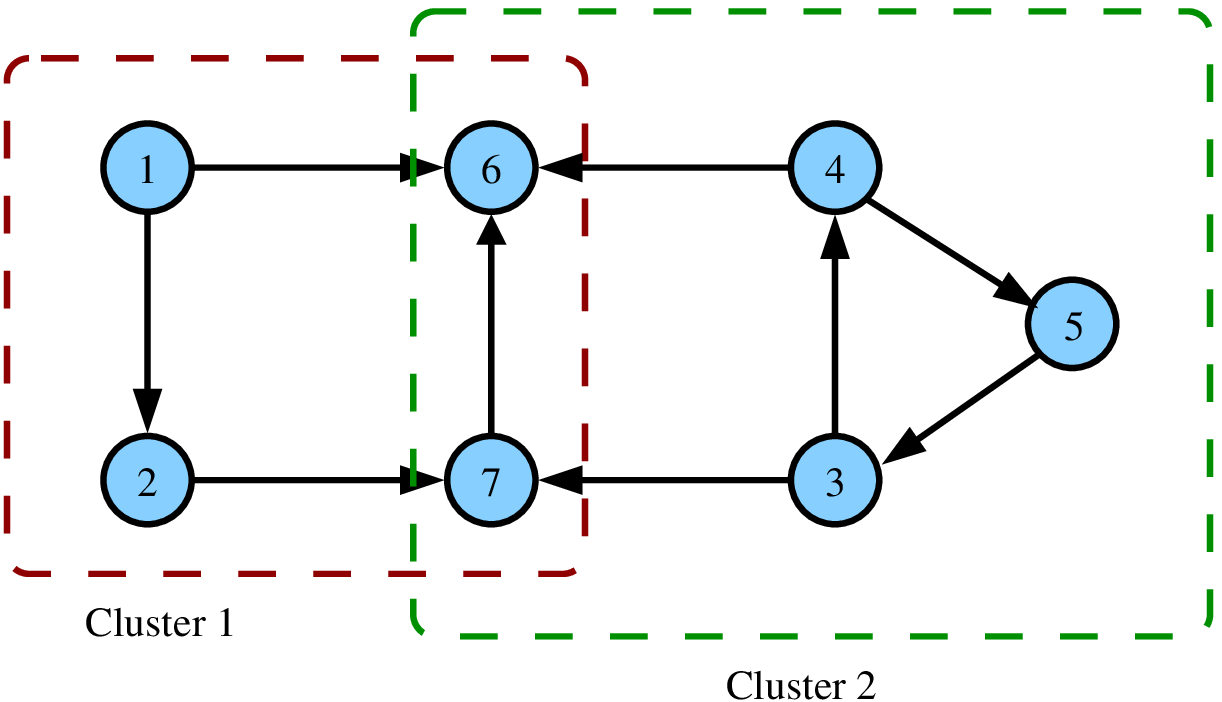}
   \label{Graph1}}\\
 \subfigure[]{\includegraphics[width=\columnwidth]{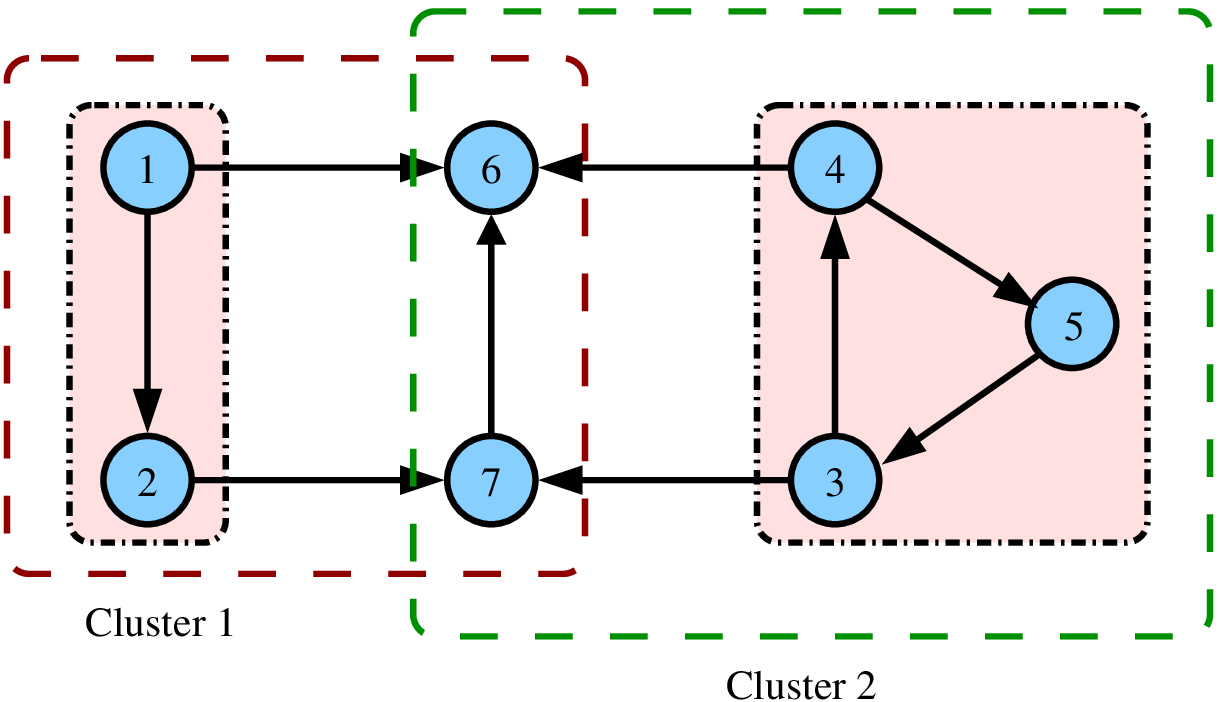}
   \label{Graph2}}
 \label{Example.Graph}
 \caption{An example of a directed graph consisting of two clusters (shown
   in Figure~(a)), with inner subgraphs shown in Figure~(b) (inside
   the shaded boxes).}
\end{figure}

An example illustrating this definition is shown in~Figure~\ref{Graph1}.
It follows from Definition~\ref{cluster} that clusters have no outgoing
connections to outside nodes. 
It also follows from Definition~\ref{cluster}, that the nodes within a
cluster which are roots of its spanning trees are not reachable from
outside the cluster. These facts are now formally stated.

\begin{proposition}\label{cluster.closed}
Consider a cluster $\mathbf{G}(i_1,\ldots,i_s)\subset \mathbf{G}$. If
$j\not\in \{i_1,\ldots,i_s\}$, then it is not reachable from
$\{i_1,\ldots,i_s\}$.  
\end{proposition}

\begin{proposition}\label{roots.unreach}
Let $i_1$ be the root node of a tree graph spanning a cluster
$\mathbf{G}(i_1,\ldots,i_s)\subset \mathbf{G}$. If $j\not \in
\{i_1,\ldots,i_s\}$, then $(j,i_1)\not\in\mathbf{E}$. 
\end{proposition}

It follows from Proposition~\ref{cluster.closed} that clusters are
the largest subgraphs reachable from within themselves, i.e., they
are reaches in the terminology of~\cite{CV-2006}. Here we call these subgraphs
clusters, to acknowledge that our formal definition is 
different; unlike~\cite{CV-2006} it involves spanning trees. 
Also, it will be shown in the next section that these subgraphs form
smallest collectively detectable clusters within collectively detectable
networks, hence the name to reflect this. 

We note the
difference between clusters and strongly connected components of a
digraph; cf.~\cite{FFS-2010}. For example, the graph in
Figure~\ref{Graph1} consists of two 
clusters comprised of the vertex sets $\{1,2,6,7\}$ and
$\{3,4,5,6,7\}$ which have a nonempty intersection. On the other hand,
strongly connected components of a
digraph, being the largest components connected by directed paths,
cannot overlap. As another important point of difference, by definition clusters
cannot have outgoing edges connecting them to the outside nodes
(but can have incoming edges) whereas 
strongly connected components of a digraph can have such outgoing edges.
From the perspective of distributed estimation, this means that the
observers that belong to a cluster do not 
share information with observers located outside this cluster, but can receive
information from other clusters. However, there exist 
observer nodes within each cluster which do not receive such
information. Such vertices form \emph{isolated} subgraphs within
the digraph~\cite{FFS-2010}. In~\cite{FFS-2010} isolated subgraphs were
associated with irreducible (i.e., strongly connected) components of the
underlying graph. In this paper, we adopt a somewhat different definition
which does not require these
subgraphs to be strongly connected; instead it emphasizes spanning trees
within clusters. This allows to establish a natural relationship between
the clusters and their inner isolated subgraphs; see Lemma~\ref{inner.prop}
below. 

\begin{definition}\label{inner-cluster}
A subgraph $\mathbf{G}(i_1,\ldots,i_r)$ is \emph{inner}, if 
\begin{enumerate}[(i)]
\item
it contains a spanning tree; and
\item   
it is a maximal subgraph with the property that its vertices $i_1,\ldots,i_r$
are not reachable from any node $j\in\mathbf{V}\backslash 
\{i_1,\ldots,i_r\}$.
\end{enumerate}   
By definition, a single-vertex subgraph, which does not have incoming edges
is inner. 
\end{definition}

\begin{lemma}\label{inner.prop}
There is exactly one inner subgraph within any cluster of a
connected graph. 
\end{lemma}

\begin{remark}
It is easy to give an example of a graph containing a strongly connected
component which has incoming edges and therefore does not contain isolated
(inner) 
subgraphs. On the other hand, 
since each strongly connected component of a digraph contains a spanning
tree, it is a subgraph of one of the digraph's clusters. This leads to the
conclusion that the number of strongly connected components in a digraph is
greater than or equal to the number of clusters. 
\end{remark}

We now present the main results of this section.  

\begin{theorem}\label{T1}
The multiplicity of the zero eigenvalue of $\mathcal{L}$ is equal to the
number of clusters in the graph $\mathcal{G}$. 
\end{theorem}

The proof of Theorem~\ref{T1} 
relies on  Lemma~\ref{inner.prop}. According to this lemma, it suffices to
show that the 
multiplicity of the zero eigenvalue of $\mathcal{L}$ is equal to the 
number of  inner subgraphs in the graph $\mathcal{G}$. The latter proof
proceeds in a manner similar to proving a similar claim in Theorem~2
of~\cite{FFS-2010} involving strongly connected components of the graph and
its isolated subgraphs. The key observation underlying the proof is that 
by definition, for any  inner
subgraph $\mathbf{G}(i_1,\ldots,i_s)$, we have $\mathcal{L}_{ij}=0$, if
$i \in\{i_1,\ldots,i_s\}$ and $j\not \in\{i_1,\ldots,i_s\}$.   
This shows that by permuting rows and
columns of $\mathcal{L}$, the Laplacian matrix can be represented in the
following block-matrix form  
\begin{equation}
  \label{block.Laplacian}
  \mathcal{L}=\left[
    \begin{array}{cccc}
      \mathcal{L}_{11} & \ldots & 0 & 0 \\
       0 & \ddots & 0 & 0 \\
       0 & \ldots & \mathcal{L}_{ll} & 0 \\
       F_1 & \ldots & F_l & R 
    \end{array}
\right],
\end{equation}
where the first $l$ rows of blocks correspond to  inner subgraphs
of $\mathbf{G}$, and the remaining rows correspond to nodes that do not
belong to any of the  inner subgraphs. Furthermore, since by
definition  inner subgraphs do not have incoming edges, each of the
diagonal blocks $\mathcal{L}_{ii}$ is a Laplacian matrix of the
corresponding subgraph. Also, it can be shown that $R$ is nonsingular.

Theorem~\ref{T1} allows to determine the basis of the null-space of
$\mathcal{L}$. This result will be instrumental in the analysis of the
the unobservable subspace of system interconnections given in the next
section. Without loss of generality, in the following corollary we assume
that the vertices of the underlying digraph are ordered so that the Laplace
matrix $\mathcal{L}$ is of the form (\ref{block.Laplacian}). 

\begin{corollary}\label{L.eigenvectors}
$    \Ker\mathcal{L}=\mathrm{Span}(b_1,\ldots,b_l)$
where $l$ is the multiplicity of the zero eigenvalue of $\mathcal{L}$, and 
\begin{eqnarray*}
b_1&=&\left[\begin{array}{c}
   \mathbf{1}_{\dim{\mathcal{L}_{11}}} \\ 
\mathbf{0}_{\dim{\mathcal{L}_{22}}+\ldots+
\dim{\mathcal{L}_{ll}}} \\
    -R^{-1}F_i\mathbf{1}_{\dim{\mathcal{L}_{11}}} 
  \end{array}\right], \\
b_i&=&\left[\begin{array}{c}
\mathbf{0}_{\dim{\mathcal{L}_{11}}+\ldots+
\dim{\mathcal{L}_{i-1,i-1}}} \\
   \mathbf{1}_{\dim{\mathcal{L}_{ii}}} \\ 
\mathbf{0}_{\dim{\mathcal{L}_{i+1,i+1}}+\ldots+
\dim{\mathcal{L}_{ll}}} \\
    -R^{-1}F_i\mathbf{1}_{\dim{\mathcal{L}_{ii}}} 
  \end{array}\right], \quad i=2,\ldots,l.
\end{eqnarray*}
\end{corollary}
\mbox{}

\section{The distributed detectability problem}\label{main}

The property of collective detectability refers to the ability of a network
of consensus-based observers 
 \begin{eqnarray}
    \dot{\hat x}_i=A\hat x_i + L_i(y_i(t)-C_{i}\hat x_i)+K_i\sum_{j\in
      \mathbf{V}_i}H_i(\hat x_j-\hat x_i),   \label{UP7.C.d}
\\
 \hat x_i(0)=0, \nonumber
\end{eqnarray}
to provide an asymptotically accurate estimate of the state of a plant 
\begin{equation}
  \label{eq:plant}
  \dot x=Ax+B\xi(t), \quad x(0)=x_0,
\end{equation}
from their local measurements of the form 
\begin{equation}\label{U6.yi}
y_i(t)=C_{i}x(t)+D_{i}\xi(t)+{\bar D_{i}}\xi^i(t). 
\end{equation}
Here $x\in\mathbf{R}^n$ is the state of the plant, $ \hat{x}_i$ is its 
estimate calculated at node $i$, $\xi\in {R}^m$, $\xi\in {R}^m$, ${\xi^i}\in {R}^{m_i}$ represent the plant
uncertainty and the measurement uncertainty at the local sensing node $i$,
$A$, $B$, $C_i$, $D_i$, $\bar D_i$ are given constant matrices of
corresponding dimensions, and $H_i$ are given matrices
which describe the information shared by nodes $j\in\mathbf{V}_i$ with
$i$. 

The matrices $L_i$, $K_i$ are parameters of the filters.
In a distributed state estimation problem,  they are
to be determined to ensure that the size of the estimation error, $\|\hat
x_i(t)-x(t)\|$ reduces to 0 asymptotically, in an $L_2$ sense 
or remains bounded in some sense, depending on
the nature of the disturbance signals 
$\xi$, $\xi_i$. For such matrices to exist the following detectability
property is naturally expected to hold.

Define the matrices
\begin{eqnarray*}
&&\bar A=I_N\otimes A, \quad \bar H=\left[\bar
  H_{ij}\right]_{i,j=1,\ldots,N}\\
&& \mbox{where }\bar H_{ij}=\begin{cases}p_i H_i & \text{if $j=i$},\\ 
-\mathbf{a}_{ij} H_i & \text{if $j\neq i$},
\end{cases}\\
&& \bar C=\diag\left[C_1,\ldots,C_N\right]. 
\label{Cs}
\end{eqnarray*}

\begin{definition}
The system consisting of the plant (\ref{eq:plant}), the measurements
(\ref{U6.yi}) and the observers (\ref{UP7.C.d}) is said to be \emph{collectively
detectable}, if the matrix pair  
$([\bar C',\bar H']',\bar A)$ is detectable.
\end{definition} 

\begin{remark}
Detectability of the pair $([\bar C',\bar H']',\bar A)$ is necessary but
is not sufficient for the existence of the set of observer gains $K_i,L_i$
which ensure that the matrix 
$\bar A-\diag[L_1\ldots L_N]\bar C-\diag[K_1\ldots K_N]\bar H$ is
Hurwitz. Therefore, 
the collective detectability property is a necessary condition for the
consensus-based filters (\ref{UP7.C.d}) each to provide an
estimate of the plant (\ref{eq:plant}).
\end{remark}

The pair $([\bar C',\bar H']',\bar A)$ can be detectable even if
the individual pairs $(C_i,A)$ are not detectable. As was shown
in~\cite{U7b-journal} in the case where $H_i=H$,  
$i=1,\ldots,N$, for this to be true, each network node must be able
to complement its local measurements with feedback it receives from its 
neighbours trough the interconnections. In this section, the condition
$H_i=H$ is relaxed.

Recall the definitions of undetectable and unobservable subspaces of a
matrix pair $(G,F)$, $F\in \mathbf{R}^{n\times n}$, $G\in  
  \mathbf{R}^{m\times n}$. Let $\alpha_F(s)$ denote the minimal polynomial
    of $F$, i.e., the monic polynomial of least degree such that
$\alpha_F(F)=0$~\cite{Wonham}, factored as 
$\alpha_F(s)=\alpha_F^-(s)\alpha_F^+(s)$; the zeros of
$\alpha_F^-(s)$ and $\alpha_F^+(s)$ are in the open left and closed right
half-planes of the complex plane, respectively. Note that
$\Ker\alpha_F^-(F)\cap \Ker\alpha_F^+(F)=\{0\}$, and  $\Ker\alpha_F^-(F)\oplus 
\Ker\alpha_F^+(F)= \mathbf{R}^n$~\cite{Wonham}. The undetectable subspace
of $(G,F)$ is the subspace $\bigcap_{l=1}^n\Ker(GF^{l-1})\cap
\Ker\alpha_F^+(F)$, and the unobservable subspace of $(G,F)$
is the subspace $\bigcap_{l=1}^n\Ker(GF^{l-1})$~\cite{CD-1991}.       

Define the observability matrices associated with the matrix pairs
$(C_i,A)$ and $(H_i,A)$:
\[
O_{C_i}\triangleq \left[\begin{array}{c} C_i \\ C_i A \\ \vdots \\
    C_i A^{n-1}
  \end{array}\right], \quad 
O_{H_i}=\left[\begin{array}{c} H_i \\ H_i A \\
     \vdots \\ H_i A^{n-1}
\end{array}\right].
\]
Also, consider the undetectable subspace of $(C_i, A)$
and the unobservable subspace of $(H_i,A)$, which
will be denoted $\mathcal{C}_i$, $\mathcal{O}_{H_i}$. Furthermore, let
$\bar{\mathcal{O}}$ denote the unobservable subspace of $(\bar H, \bar A)$,  
\[
\bar{\mathcal{O}}\triangleq \bigcap_{l=1}^{nN}\Ker(\bar H\bar A^{l-1}). 
\]

\begin{lemma}\label{lemma.barO}
The pair $([\bar C',\bar H']',\bar A)$ is detectable if and only if 
\begin{eqnarray}
\Ker\Big(\diag[O_{H_1},~\ldots,O_{H_N}](\mathcal{L}\otimes I_n)\Big)\cap \prod_{i=1}^N\mathcal{C}_i =\{0\}.
\label{undetect.C.1}
\end{eqnarray}
\end{lemma}

The following necessary condition for collective detectability has been
obtained in~\cite{U7b-journal} for the special case where $H_i=H$;
see~\cite[Theorem 3]{U7b-journal}. Using Lemma~\ref{lemma.barO}, this
requirement can be relaxed.  

\begin{theorem}\label{detect.theorem}
Suppose the pair $([\bar C',\bar H']',\bar A)$ is detectable. Then, for
every cluster $\mathbf{G}(i_1,\ldots,i_s)$ 
the following statements hold:
\begin{enumerate}[(i)]
\item
$\bigcap_{i\in\{i_1,\ldots,i_s\}} \mathcal{C}_i=\{0\}$;
\item for all $i\in\{i_1,\ldots,i_s\}$,
\[
\left(\bigcap_{j: i\in \mathbf{V}_j}\mathcal{O}_{H_j}\right)\cap
\mathcal{O}_{H_i} \cap \mathcal{C}_i=\{0\}.
\]
\end{enumerate}
\end{theorem}

The interpretation of claims (i) and (ii) of Theorem~\ref{detect.theorem} 
is as follows. Claim (i) states that every state of a collectively
detectable plant is necessarily detectable by at least one observer
within each cluster of the network. 
Also, condition (ii) states that communications between the observer
nodes in a collectively detectable system must be designed so that 
each plant state $x$ has at least one of the three properties at every node
of every cluster: (a) it is detectable by the node from its measurements (i.e.,
$x\not\in\mathcal{C}_i$), or (b) it is observable from the information the node
receives from its neighbours (i.e., $x\not\in\mathcal{O}_{H_i}$), or (c) 
it is observable by at least one of the neighbours with whom the node
communicates (i.e, there exists $j$ such that $i\in V_j$ and
$x\not\in\mathcal{O}_{H_j}$). In the case where $H_i=H$ $\forall i$, we
recover claim (ii) of ~\cite[Theorem 3]{U7b-journal}: for all
$i\in\{i_1,\ldots,i_s\}$, 
$\mathcal{O}_H \cap \mathcal{C}_i=\{0\}$.

Next, a sufficient condition for collective detectability is presented which
extends the corresponding condition in~\cite[Theorem~4]{U7b-journal} to the
case where the interconnection protocol matrices $H_i$ are not required to be
identical. The proof of this result relies on
Corollary~\ref{L.eigenvectors}, therefore we again assume 
that the vertices of the underlying digraph are ordered so that the Laplace
matrix $\mathcal{L}$ has the block structure (\ref{block.Laplacian}).

\begin{theorem}\label{detect.theorem.sufficient}
Suppose all the pairs $(H_i,A)$ are observable. If every cluster in the
network satisfies condition (i) of Theorem~\ref{detect.theorem}, then the
pair $([\bar C',\bar H']',\bar A)$ is detectable.  
\end{theorem}

\begin{remark}
  The result of Theorem~\ref{detect.theorem.sufficient} remains
  valid in the case where $H_i=0$ for some of the nodes. Each such node
  represents the root node in the corresponding cluster, and also belongs to
  the inner subgraph of the cluster. 
  Provided the remaining nodes have observable $(H_j,A)$, it is still true
  that $\bar{\mathcal{O}}=\Ker\mathcal{L}\otimes \mathbf{R}^n$. Hence the
  statement of the theorem holds in this case.
\end{remark}

\section{Conclusions}\label{Concl}
In this paper, we obtained necessary and sufficient conditions for
distributed detectability of a linear plant via a network of state
estimators, which were previously obtained under condition that all
observers utilize the same matrix for communication. Our results show that
in a collectively detectable system, each state in the plants' phase space
must be detectable by every observer cluster spanned by a maximal
tree. Furthermore, at every node of the network, every undetectable state
of $(C_i,A)$ must be observable through interconnections or must be
transmitted to a neighbour who can observe it. Thus, the results of this
paper elucidate the relationship between the network topology and
detectability properties of the plant and observers. In particular, the paper
makes explicit the role of spanning trees in ensuring collective
detectability.    

\bibliographystyle{plain} 
\bibliography{Val,irpnew}

\end{document}